\documentclass[symmetry,article,accept,pdftex,moreauthors]{Definitions/mdpi} 
\firstpage{1} 
\pubvolume{15}
\issuenum{10}
\articlenumber{1877}
\pubyear{2023}
\copyrightyear{2023}
\externaleditor{Academic Editors: Jinyu Li and Calogero Vetro}
\datereceived{7 September 2023} 
\daterevised{25 September 2023} 
\dateaccepted{4 October 2023} 
\datepublished{6 October 2023} 
\hreflink{https://\linebreak doi.org/10.3390/sym15101877} 

\usepackage[para]{threeparttable}
\usepackage{tablefootnote}

\Title{Lambert $W$ Random Variables and their Applications \linebreak in Loss Modelling}

\TitleCitation{Lambert $W$ Random Variables and their Applications in Loss Modelling}

\Author{{Meelis} 
 K\"a\"arik *\orcidA{}, Anne Selart, Tuuli Puhkim and Liivika Tee}

\AuthorNames{Meelis K\"a\"arik, Anne Selart, Tuuli Puhkim and Liivika Tee}

\AuthorCitation{K\"a\"arik, {M.; Selart, A.;} 
  Puhkim, T.;  Tee, L.}

\address[1]{%
 Institute of Mathematics and  Statistics, University of Tartu, Narva mnt 18,  51009 Tartu, Estonia; \linebreak {anne.selart@ut.ee} 
 (A.S.) 
}

\corres{\hangafter=1 \hangindent=1.05em \hspace{-0.82em}Correspondence: meelis.kaarik@ut.ee}

\abstract{Several distributions and families of distributions are proposed to model skewed data, e.g.,~with skew-normal and related distributions. Lambert $W$ random variables offer an alternative approach in which, instead of constructing a new distribution, a certain transformation is proposed. 
Such an approach allows the construction of a Lambert $W$ skewed version from any distribution. Here, we choose the Lambert $W$ normal distribution as a natural starting point and include the Lambert $W$ exponential distribution due to the simplicity and shape of the exponential distribution, which, after skewing, may produce a reasonably heavy tail for loss models. In the theoretical part, we focus on the mathematical properties of obtained  distributions, including the  range of skewness.
In the practical part, the suitability of the corresponding Lambert $W$ transformed distributions is evaluated on real insurance data. Finally, the results are compared with those obtained using common loss distributions.}

\keyword{asymmetry; skewness; loss distributions; non-life insurance; probability distributions; Lambert $W$ function} 

\begin{document}


\section{Introduction}

{Loss modelling} 
 is an essential part of actuarial and financial mathematics. Several distributional models have been applied over the years, and~the increasing volumes of data and computational power have motivated the use of even more complex distributions to fit the~data. 

In the actuarial and financial fields, the data are usually skewed. Several classical distributions can be used to fit skewed data (see, e.g.,~\citep{hogg1984,klugman2012loss}). A generic approach for skewing symmetric distributions was introduced in \citet{Azzalini1985}, where the shape of the normal distribution is deformed by a certain skewness parameter. Similarly, other asymmetric distributions (e.g., skew $t$-distribution) have been developed \citep{AzzaliniCapitanio2003}. Unified overviews of skewed distributions are provided in~\citep{genton2004skew,nadarajah2003}, while a review of different applications of skew-elliptical distributions in actuarial and financial mathematics is provided in~\citep{Adcock2015}. 

In \citep{Goerg2011}, another method of generating skewness {was} introduced through the Lambert $W$ function that, when applied to symmetric distributions, can produce skewness and a heavy tail. In~addition, Lambert $W$ random variables can be seen as a generalization, as the input distribution can be arbitrary and not necessarily symmetric. When using the Lambert $W$ function, instead of using the parametric manipulation of the original symmetric density function to introduce skewness, the~random variable itself is transformed.

{Another Lambert \textit{W} transformation related to random variables was studied in \citep{witkovsky2014}, namely, a class of log-Lambert \textit{W} random variables with applications to likelihood-based inference of normal random variables.}

{A different approach using the Lambert \textit{W} function was introduced in \citep{iriarte2020,iriarte2021}, where the transformation is applied to the cumulative distribution function of the continuous positive valued random variable.}

The Lambert \textit{W} function has proven useful in mathematics, physics, chemistry, biology, engineering, risk theory, and other fields, though it has been less widely used in statistical modelling. 
{Nonetheless, there are a number of noteworthy examples. In \citep{peterson2021}, the~Lambert $W$ approach was applied to normalize a vector regardless of its actual distribution. The~use of the Lambert $W$ distribution in matrix factorization with an implementation in probabilistic programming was presented in \citep{klami2019lambert}.}
The Lambert $W$ function has been used to derive the exact distribution of the likelihood ratio test statistic and to solve related problems in \citep{stehlik2003,stehlik2006,stehlik2014}. 

The approach of modelling the skewed random variables and symmetrizing the data using the Lambert $W$ function as a variable transformation was used in \citep{Goerg2011,goerg2015, Goerg2022pakett, peterson2021}. We use \citep{Goerg2011} as the basis of our construction in this~paper.

The rest of this paper is organized as follows. In~the first section, we provide a short overview of the Lambert $W$ function. In~Section~\ref{sec:LambertWrv}, general definitions and the expressions of the cumulative density functions and probability density functions of the Lambert $W$ random variables are introduced, followed by more detailed results concerning the Lambert $W$ normal and exponential distributions. {In~Section~\ref{sec:LambertWapp},} 
 we describe the results of fitting the Lambert $W$ normal and exponential distributions to two insurance-related datasets, then compare the fit with {several} typical insurance models. Proofs of several properties, technical details of estimation, and~additional figures showing the fitted distributions are presented in the~{Appendices}~\ref{sec:app_proofs}--\ref{sec:app_joonised}.  

\section{The Lambert $W$ Function and Its~Properties}
In the following, we define the Lambert $W$ function and provide a brief overview of its properties; refer to \citep{Brito2008, Dence2013, Corless1996} for more details on the~topic.

The Lambert $W$ function is a set of inverse functions for the following function: $f(x')=x'e^{x'}$ $(x' \in \mathbb{R})$, in~other words,

\begin{equation*}
\label{eq:Wintro}
x'=f^{-1}(x'e^{x'})=W(x'e^{x'}).
\end{equation*}
{Substituting} 
 $x = x'e^{x'}$ leads to the definition of the Lambert $W$ function.

\begin{Definition}
The Lambert $W$ function $W(x)$ is defined by the following equality:
\begin{equation}
\label{eq:Wfun}
W(x)e^{W(x)}=x, \quad x \in \left[-\frac{1}{e},\infty\right).
\end{equation}
\end{Definition}

Note that, in~general, the~function $W(x)$ can be defined for real or complex arguments, and~that Equation \eqref{eq:Wfun} has infinitely many solutions, most of which are complex. Following the notation of \citep{Corless1996}, we denote the different branches of the function by $W_{k}(x)$, where the branch index $k \in \{0,\pm 1, \pm 2, \dots\}$ and $x \in \mathbb{C}$. 
For real $x$, all branches other than $W_{0}(x)$ and $W_{-1}(x)$ are complex. For~$x \in \left(-\infty,-\frac{1}{e}\right)$, the~equation has only complex solutions.
We denote the branch corresponding to $W(x) \geq -1$ by $W_{0}(x)$, which we call the principal branch, and~the branch corresponding to $W(x) \leq -1$ by $W_{-1}(x)$, which we call the non-principal~branch.

\begin{figure}[H]
\includegraphics[width=0.6\textwidth]{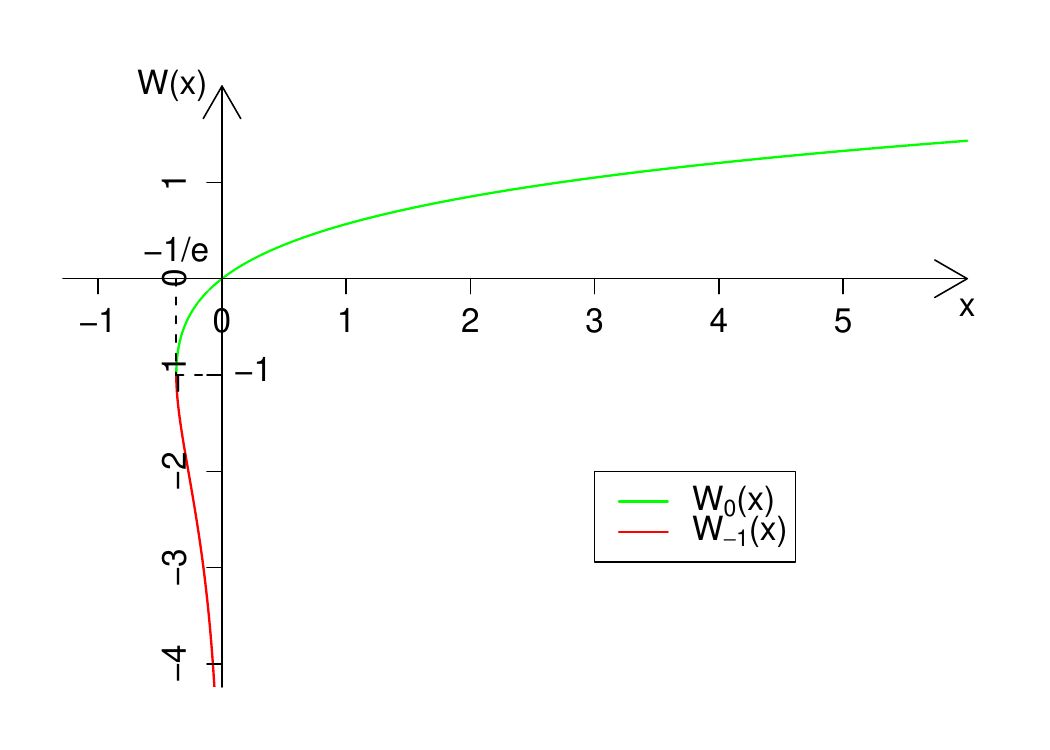}
\caption{{Lambert} 
 $W$ function.}\label{fig:Wfunc}
\end{figure}

Among the characteristic properties of the function (see Figure~\ref{fig:Wfunc} as well) are:
\begin{enumerate}
\item $W(0)=0$
\item $W_{0}\left(-\frac{1}{e}\right)=W_{-1}\left(-\frac{1}{e}\right)=-1$
\item $W(e)=1$
\item $W(1)=e^{-W(1)}=\ln\left(\frac{1}{W(1)}\right)=-\ln W(1) \approx 0.5671433$
\item $\lim_{x \rightarrow 0-} W_{-1}(x) = -\infty$
\item $\lim_{x \rightarrow \infty} W_{0}(x) = \infty$
\end{enumerate}

Based on its construction as an inverse of a certain exponential function, the~asymptotes of $W$ are similar to those of the natural logarithm. More precisely, the limits can be found as follows:

\begin{equation*}
\lim_{x \rightarrow \infty}\frac{W_{0}(x)}{\ln{x}}= \lim_{x \rightarrow \infty}\frac{xW_{0}(x)}{x(1+W_{0}(x))}= \lim_{x \rightarrow \infty}\frac{1}{\frac{1}{W_{0}(x)}+1}=1
\end{equation*}
and
\begin{equation*}
\lim_{x \rightarrow 0-}\frac{W_{-1}(x)}{\ln{(-x)}}= \lim_{x \rightarrow 0-}\frac{xW_{-1}(x)}{x(1+W_{-1}(x))}= \lim_{x \rightarrow 0-}\frac{1}{\frac{1}{W_{-1}(x)}+1}=1.
\end{equation*}
{At the same} 
 time, the~absolute difference between the Lambert's $W$ function and the natural logarithm $\vert W_{0}(x)-\ln{x} \vert$ goes to infinity for $x \rightarrow \infty$ \citep{Dence2013}.

\section{Lambert $W$ Random~Variables}\label{sec:LambertWrv}


\subsection{Definitions}
Next, we present the definitions of different types of Lambert random variables based on \citet{Goerg2011}. We provide the formulae of the cumulative distribution function (cdf) and probability density function (pdf) for scale and location--scale random variables.

\begin{Definition}
Let $U$ be a continuous random variable with a cdf $F_{U}(u)=\mathbb{P}(U \leq u)$, $u \in \mathbb{R}$ and pdf $f_{U}(u)$; then,
\begin{equation}
\label{eq:LambertW-ncns}
Y:=U\exp(\gamma U), \quad \gamma \in \mathbb{R}
\end{equation}
is a noncentral and nonscaled Lambert $W \times F_{U}$ random variable with skewness parameter $\gamma$.
\end{Definition}

The skewness parameter $\gamma$ can take any value on the real line; however,~as the exponential function is always positive, the~transformation \eqref{eq:LambertW-ncns} preserves the sign. Thus, if~$\gamma=0$, then $Y=U$. The~effect of the transformation on the shape of the distribution depends on the original variable $U$.
If $U$ has both positive and negative values, then positive $\gamma$ folds back the tail with negative values at a point $-\frac{1}{\gamma}$, relocating part of  negative $U$ values, while~on the positive side the values move further away, making the right tail heavier. Negative $\gamma$ acts the other way around. Note that for a skewed $U$, the~Lambert $W$ transform can produce a more symmetric random~variable.

The transformation in \eqref{eq:LambertW-ncns} is not scale- or location-invariant. In order to~keep these properties, {which are needed, for~example, to~construct the Lambert $W$ normal random variables}, it is necessary to include the transformed variable's location and scale parameters in the definition. For more details about the location-scale family of distributions, refer to~\mbox{\citep{casella2002}~(pp.~116--121).}

\begin{Definition} 
\label{def:location-scale}
Let  $X$ be a continuous random variable from a location-scale family with cdf $F_{X}(x\vert\boldsymbol{\beta})$, where $\boldsymbol{\beta}$ is the corresponding parameter vector. 
Let $\displaystyle U=\frac{X-\mu}{\sigma}$ be the zero-mean unit variance version of $X$. Then,
\begin{equation}
\label{eq:asukoha-skaala}
Y:=\{U\exp(\gamma U)\}\sigma+\mu, \quad \gamma \in \mathbb{R}, \enskip \sigma>0
\end{equation}
is a location-scale Lambert $W \times F_X$ random variable with parameter vector $(\boldsymbol{\beta},\gamma)$.
\end{Definition}

If $\gamma >0$, the~location-scale  Lambert  $W \times F_X$ random variable takes values in the interval $(\mu- \frac{\sigma}{\gamma e}, \infty)$. For~{a} negative $\gamma$, on~the contrary, $Y$ has an upper bound, and the~values are in the interval $(-\infty, \mu - \frac{\sigma}{\gamma e})$.
 
For $\gamma>0$, the~cdf and pdf of a location-scale Lambert $W\times F_X$ random variable are respectively
\begin{equation}
\label{eq:loc-scale-cdf}
F_{Y}(y\vert\boldsymbol{\beta},\gamma)=
\begin{cases}
0,\quad&\text{if $y\leq \mu-\frac{\sigma}{\gamma e}$} ,\\
F_{X}\left(\left.\frac{W_{0}(\gamma z)}{\gamma}\sigma+\mu\right\vert\boldsymbol{\beta}\right) \\ \quad \quad \quad
-F_{X}\left(\left.\frac{W_{-1}(\gamma z)}{\gamma}\sigma+\mu\right\vert\boldsymbol{\beta}\right), \quad&\text{if $\mu-\frac{\sigma}{\gamma e} < y <\mu$} ,\\
F_{X}\left(\left.\frac{W_{0}(\gamma z)}{\gamma}\sigma+\mu\right\vert\boldsymbol{\beta}\right), \quad&\text{if $y \geq \mu$},
\end{cases}
\end{equation}
and
\begin{equation}
\label{eq:loc-scale-pdf}
f_{Y}(y\vert\boldsymbol{\beta},\gamma)=
\begin{cases}
0,\quad&\text{if $y\leq\mu-\frac{\sigma}{\gamma e}$},\\
f_{X}\left(\left.\frac{W_{0}(\gamma z)}{\gamma}\sigma+\mu\right\vert\boldsymbol{\beta}\right)\frac{W'_{0}(\gamma z)}{\gamma}\\ \quad \quad \quad-f_{X}\left(\left.\frac{W_{-1}(\gamma z)}{\gamma}\sigma+\mu\right\vert\boldsymbol{\beta}\right)\frac{W'_{-1}(\gamma z)}{\gamma},
\quad&\text{if $\mu-\frac{\sigma}{\gamma e} < y <\mu$} ,\\
f_{X}\left(\left.\frac{W_{0}(\gamma z)}{\gamma}\sigma+\mu\right\vert\boldsymbol{\beta}\right)\frac{W'_{0}(\gamma z)}{\gamma}, \quad&\text{if $y \geq \mu$},
\end{cases}
\end{equation}
where $\displaystyle z=\frac{y-\mu}{\sigma}$ and we denote the derivative of $W(\gamma z)$ by $z$ as
\begin{equation}
\label{eq:Wprim}
W'(\gamma z) = \frac{dW(\gamma z)}{dz}=\frac{\exp({-W(\gamma z}))} {1+W(\gamma z)}\gamma =\frac{W(\gamma z)}{ z(1+W(\gamma z))}.
\end{equation}
{In} 
 \eqref{eq:Wprim}, the~principal and non-principal branches are not distinguished, as the same holds for~both.

The derivation of these expressions can be found in \citet{Goerg2011}. The~derivation and resulting expressions for~$\gamma<0$ are similar, except that~the three regions considered are pivoted: the first region is $y\leq \mu$, where only the principal branch is used; the~second region is $\mu<y< \mu-\frac{\sigma}{\gamma e}$, where both branches are used; and $y\geq\mu-\frac{\sigma}{\gamma e}$~for last region, where the cdf reaches 1 and~the pdf is equal to~0. 

For a non-negative $X$ from the scale family, {for example, an exponentially-distributed $X$}, we can define the corresponding scale-family Lambert random variable as follows.

\begin{Definition} 
\label{def:scale}
Let  $X$ be a non-negative continuous random variable from a scale family with cdf $F_{X}(x\vert\boldsymbol{\beta})$, where $\boldsymbol{\beta}$ is the parameter vector. Let $\displaystyle U=\frac{X}{\sigma}$ be the unit-variance version of $X$. Then,
\begin{equation}
\label{eq:skaala}
Y:=\{U\exp(\gamma U)\}\sigma = X\exp(\gamma X/\sigma), \quad \gamma \in \mathbb{R}, \enskip \sigma>0
\end{equation}
is a scale Lambert $W \times F_X$ random variable with parameter vector $(\boldsymbol{\beta},\gamma)$.
\end{Definition}

If $\gamma>0$, then the~cdf and pdf for a scale Lambert random variable can be found easily, as the transformation \eqref{eq:skaala} takes values only on the positive side of the real line; as~we apply the transformation $W$ on positive arguments {as well}, only the {principal}  branch plays a role. Hence, the~cdf has the following form:
\begin{equation}
\label{eq:scale-cdf}
F_{Y}(y\vert\boldsymbol{\beta},\gamma)=
\begin{cases}
    0, & \textup{if } y < 0,\\
    F_{X}\left(\left.\frac{W_0(\gamma y/\sigma)}{\gamma} \sigma\right\vert\boldsymbol{\beta}\right), & \textup{if } y \geq 0.
\end{cases}
\end{equation}
{Taking} 
 the derivative of \eqref{eq:scale-cdf}, we obtain the following form for the pdf:
\begin{equation}
\label{eq:scale-pdf}
f_{Y}(y\vert\boldsymbol{\beta},\gamma)=
\begin{cases}
    0, & \textup{if } y < 0,\\
     f_{X}\left(\left.\frac{W_0(\gamma y/\sigma)}{\gamma} \sigma\right\vert\boldsymbol{\beta}\right)\frac{\exp({-W_0(\gamma y/\sigma)})}{1+W_0(\gamma y/\sigma)} & \textup{if } y \geq 0.
\end{cases}
\end{equation}

{Our} 
 primary focus is on positive $\gamma$ that produces a heavier right tail to right-skewed distribution, possibly making the distribution  more suitable for describing insurance losses. Yet, the~results for $\gamma<0$ are not as straightforward as for the location-scale family case. Thus, to~complete the theory, we analyze this situation as well and derive the cdf and pdf.
First, the~cdf:
\begin{equation*}
\begin{aligned}
F_{Y}(y)&=\mathbb{P}(Y\leq y)=\mathbb{P}(U\exp({\gamma U})\sigma\leq y) 
=\mathbb{P}(\gamma U\exp({\gamma U})\geq \gamma y/\sigma) \\
& = 1 - \mathbb{P}(\gamma U\exp{(\gamma U)}\leq \gamma y/\sigma).
 \end{aligned}   
\end{equation*}
{Now,} 
 as~the argument $\gamma y/\sigma$ is negative for $y>0$, both branches are needed when we apply the Lambert function.~Hence,
\begin{equation*}
\begin{aligned}
F_{Y}(y)&= 1- \mathbb{P}(W_{-1}(\gamma y/\sigma)\leq \gamma U \leq W_{0}(\gamma y/\sigma))\\
&= 1- \mathbb{P}(W_{-1}(\gamma y/\sigma)/\gamma\geq  U \geq W_{0}(\gamma y/\sigma)/\gamma)\\
&= 1- 
F_X\left(\left.\frac{W_{-1}(\gamma y/\sigma)}{\gamma}\sigma\right\vert\boldsymbol{\beta}\right)   + 
F_X\left(\left.\frac{W_{0} (\gamma y/\sigma)}{\gamma}\sigma\right\vert\boldsymbol{\beta}\right).
 \end{aligned}   
\end{equation*}
{The} 
principal and non-principal branches are equal at point $y=-\frac{\sigma}{\gamma e}$; thus, this is the point where the cdf $F_Y$ reaches 1.
In summary, if~$\gamma<0$, then
\begin{equation}
\label{eq:scale-cdf2}
F_{Y}(y\vert\boldsymbol{\beta},\gamma)=
\begin{cases}
0,\quad&\text{if $y\leq 0$} ,\\
1- F_{X}\left(\left.\frac{W_{-1}(\gamma y/\sigma)}{\gamma}\sigma\right\vert\boldsymbol{\beta}\right) 
+F_{X}\left(\left.\frac{W_{0}(\gamma y/\sigma)}{\gamma}\sigma\right\vert\boldsymbol{\beta}\right), \quad&
\text{if  $0 < y <-\frac{\sigma}{\gamma e}$} ,\\
1, \quad&\text{if  $y \geq -\frac{\sigma}{\gamma e}$}
\end{cases}
\end{equation}
and the corresponding pdf is
\begin{equation}
\label{eq:scale-pdf2}
f_{Y}(y\vert\boldsymbol{\beta},\gamma)=
\begin{cases}
0,\quad & \text{if $y\leq 0$ or $y \geq -\frac{\sigma}{\gamma e}$},\\
 f_{X}\left(\left.\frac{W_{0} (\gamma y/\sigma)}{\gamma}\sigma\right\vert\boldsymbol{\beta}\right)\frac{\exp({-W_0(\gamma y/\sigma)})}{1+W_0(\gamma y/\sigma)}\\ \quad \quad \quad
-f_{X}\left(\left.\frac{W_{-1}(\gamma y/\sigma)}{\gamma}\sigma\right\vert\boldsymbol{\beta}\right)\frac{\exp({-W_{-1}(\gamma y/\sigma)})}{1+W_0(\gamma y/\sigma)},
&\text{if  $0 < y <-\frac{\sigma}{\gamma e}$}.
\end{cases}
\end{equation}


\subsection{Lambert $W$ Normal~Distribution}\label{sec:LambertWN}

In this section, we apply the Lambert location-scale transformation \eqref{eq:asukoha-skaala} on a normal random variable $X\sim N(\mu, \sigma)$. The~resulting random~variable

\begin{equation*}
    Y = \frac{X-\mu}{\sigma} \exp{\left(\gamma \frac{X-\mu}{\sigma} \right)}\sigma + \mu
\end{equation*}
is a Lambert $W\times N(\mu, \sigma)$ random variable with parameter vector $(\mu, \sigma, \gamma)$.
Without loss of generality, we assume that the skewness parameter $\gamma$ is positive; the situation is mirrored for negative $\gamma$-s (i.e., left skew instead of right skew). Using \eqref{eq:loc-scale-cdf}, the~cdf for a positive skewness parameter $\gamma$ can be written~as
\begin{equation*}
F_{Y}(y\vert \mu,\sigma,\gamma)=
\begin{cases}
0,\quad&\text{if $y\leq \mu -\frac{\sigma}{\gamma e}$} ,\\
\Phi\left(\frac{W_{0}(\gamma z)}{\gamma}\right)
-\Phi\left(\frac{W_{-1}(\gamma z)}{\gamma}\right), \quad&\text{if $\mu -\frac{\sigma}{\gamma e} < y <\mu$},\\
\Phi\left(\frac{W_{0}(\gamma z)}{\gamma}\right), \quad&\text{if $y \geq \mu$},
\end{cases}
\end{equation*}
where $z = \frac{y-\mu}{\sigma}$ and $\Phi$ is the standard normal cdf. Likewise, using \eqref{eq:loc-scale-pdf}, we obtain the pdf for~$\gamma>0$ as
\begin{equation*}
\label{eq:W_N_pdf1}
f_{Y}(y\vert \mu,\sigma,\gamma)=
\begin{cases}
0,\quad&\text{if  $y\leq\mu -\frac{\sigma}{\gamma e}$},\\
f_0\left(\frac{y-\mu}{\sigma}\right)-f_{-1}\left(\frac{y-\mu}{\sigma}\right),\quad&\text{if $\mu -\frac{\sigma}{\gamma e} < y <\mu$} ,\\
f_0\left(\frac{y-\mu}{\sigma}\right), \quad&\text{if $y \geq \mu$},
\end{cases}
\end{equation*}
where $f_0(z)$ and $f_{-1}(z)$ are the components of the pdf corresponding to the principal and non-principal branch, respectively:
\begin{align}
    f_0(z) &= \frac{1}{\sqrt{2\pi}}\exp\left({-\frac{(W_{0}(\gamma z))^2}{2\gamma^2}}\right)\frac{\exp({-W_{0}(\gamma z)})} {1+W_{0}(\gamma z)}, \label{eq:f0}\\
    f_{-1}(z) &=\frac{1}{\sqrt{2\pi}}\exp\left({-\frac{(W_{-1}(\gamma z))^2}{2\gamma^2}}\right)\frac{\exp({-W_{-1}(\gamma z)})} {1+W_{-1}(\gamma z)}.\label{eq:fm1}
\end{align}

\begin{figure}[H]
\includegraphics[width=0.75\textwidth]{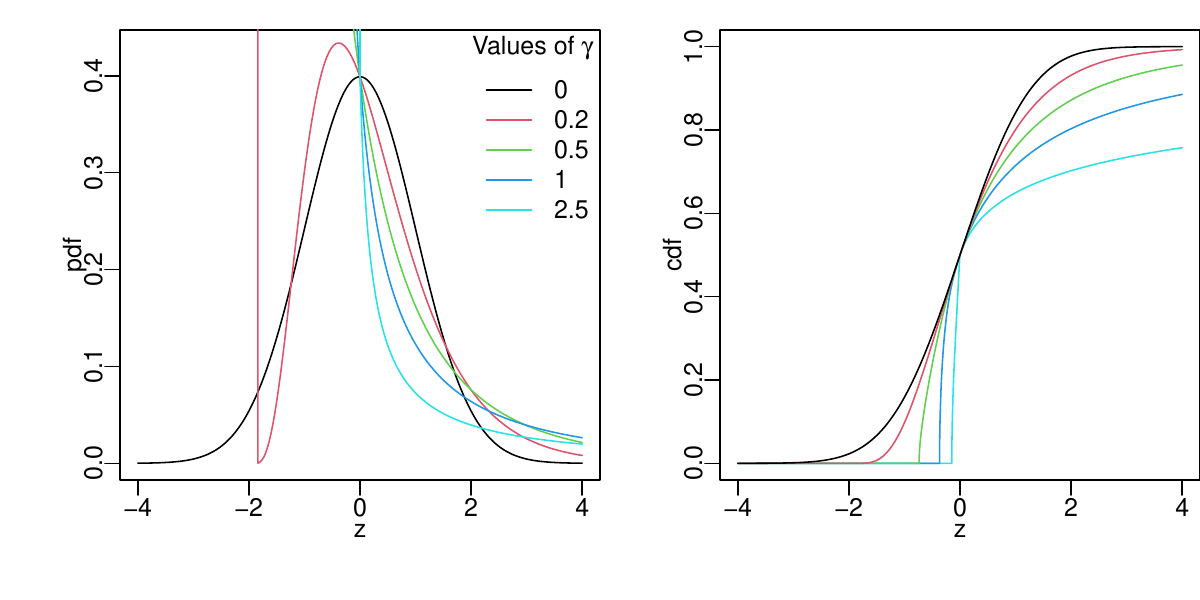}
\caption{Plots of the pdf (left panel) and cdf (right panel) of $W\times N(0, 1)$ distributions with different $\gamma$~values.}
\label{fig:WNorm-pdf-cdf}
\end{figure}

{Examples} 
  of the cdf and pdf for the Lambert $W\times N(0,1)$ distribution with $\gamma>0$ are shown in Figure~\ref{fig:WNorm-pdf-cdf}.
 \vspace{-6pt}

In the following, we provide several results that describe the behaviour of the pdf of a Lambert $W$ normal random variable. To~keep our proofs technically cleaner, the~analysis is applied to Lambert $W\times N(0,1)$ random variables, as generalization to Lambert $W\times N(\mu,\sigma)$ is straightforward. Proofs of these lemmas are presented in Appendix~\ref{sec:app_proofs}.

\begin{Lemma} The pdf  of a Lambert $W\times N(0,1)$ random variable $Z$, $f_Z$ has an asymptote at $-\frac{1}{\gamma e}$: \[\lim_{z\to-\frac{1}{\gamma e}} f_Z(z) = \infty.\]
\end{Lemma}

\begin{figure}[H]
\includegraphics[width=0.75\textwidth]{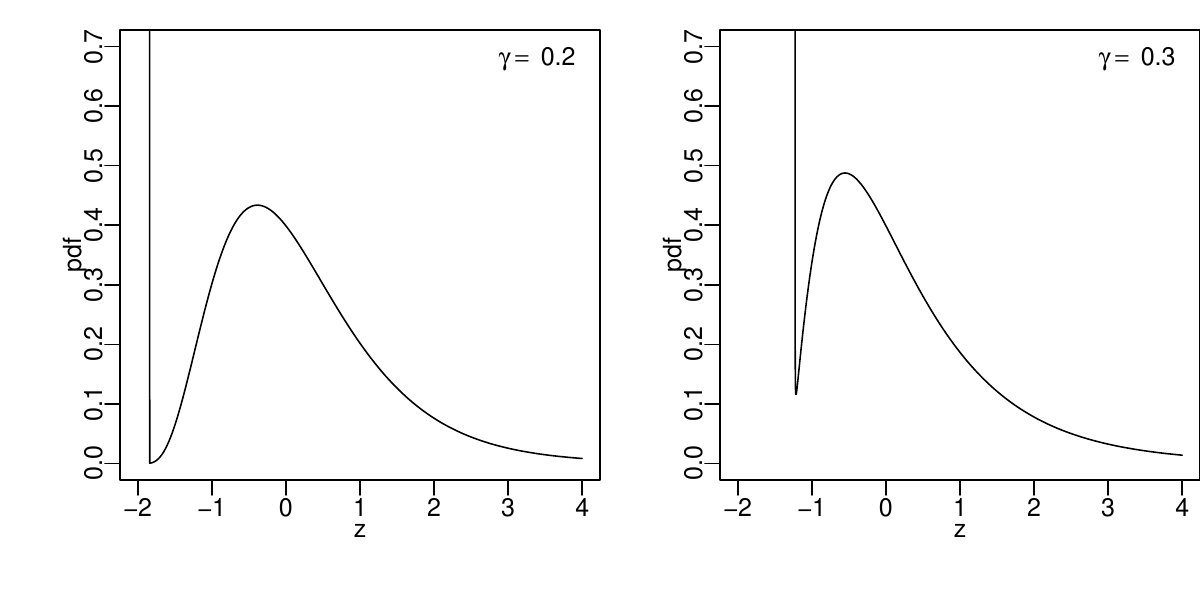}
\caption{Examples of Lambert $W\times N(0, 1)$ pdf with $\gamma = 0.2$ and $\gamma = 0.3$. }
\label{fig:WNorm-gamma0203}
\end{figure}

\begin{figure}[H]
\includegraphics[width=0.8\textwidth]{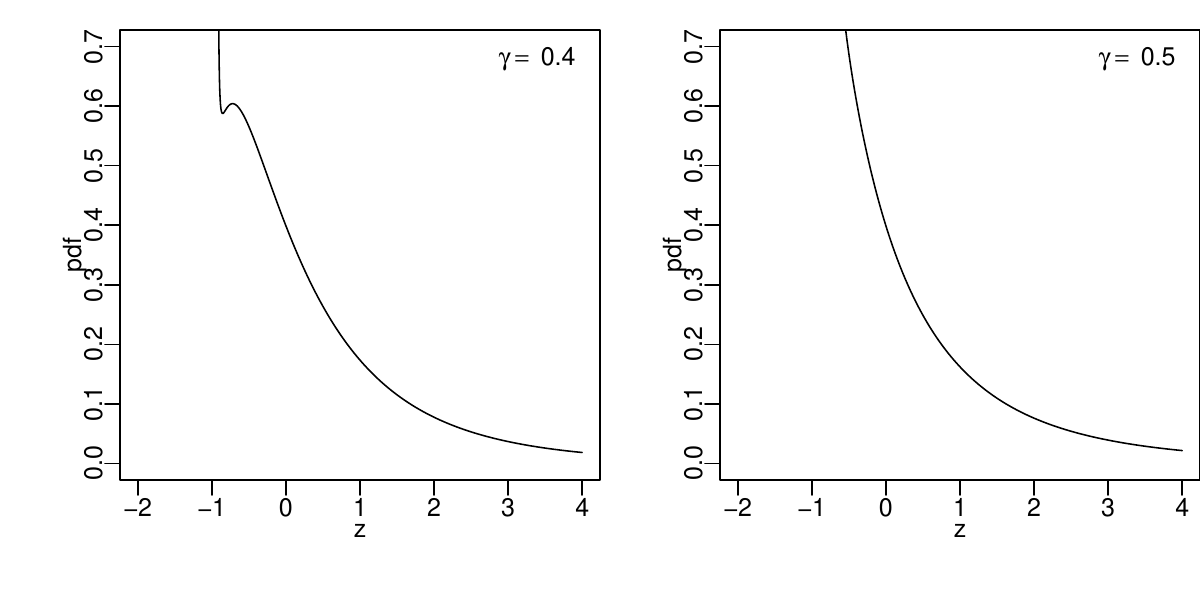}
\caption{Examples of Lambert $W\times N(0, 1)$ pdf with $\gamma = 0.4$ and $\gamma = 0.5$. }
\label{fig:WNorm-gamma0405}
\end{figure}

\begin{figure}[H]
\includegraphics[width=0.81\textwidth]{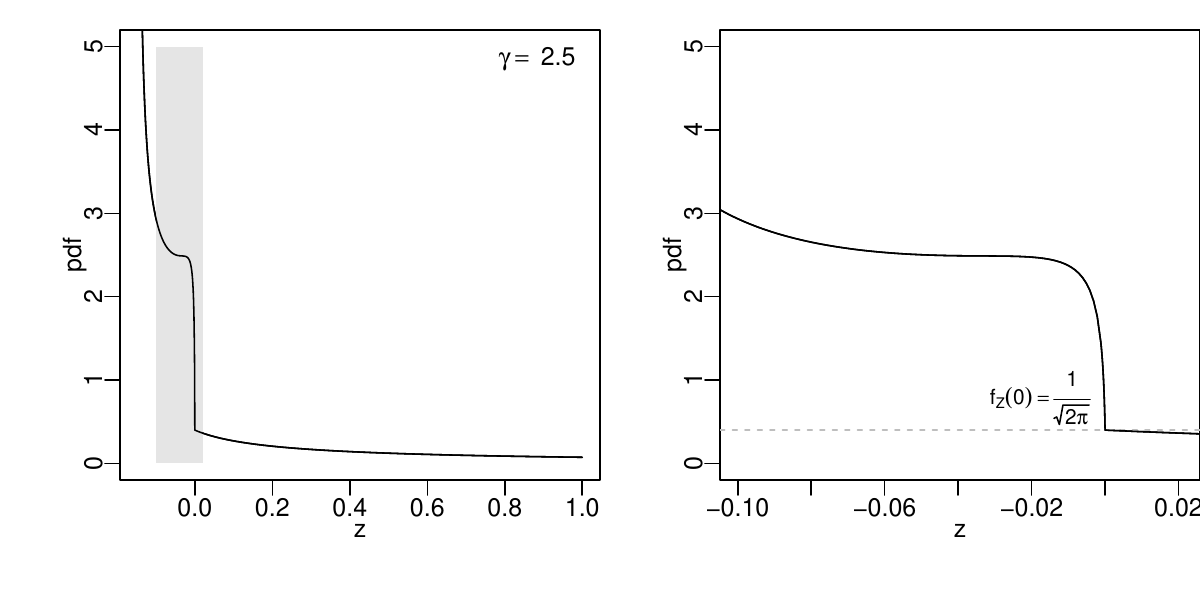}
\caption{{Example} 
 of Lambert $W\times N(0, 1)$ pdf when $\gamma = 2.5$. The~right panel shows a closer view of the interval marked with grey in the left~panel.}
\label{fig:WNorm-gamma25}
\end{figure}

\begin{figure}[H]
\includegraphics[width=0.81\textwidth]{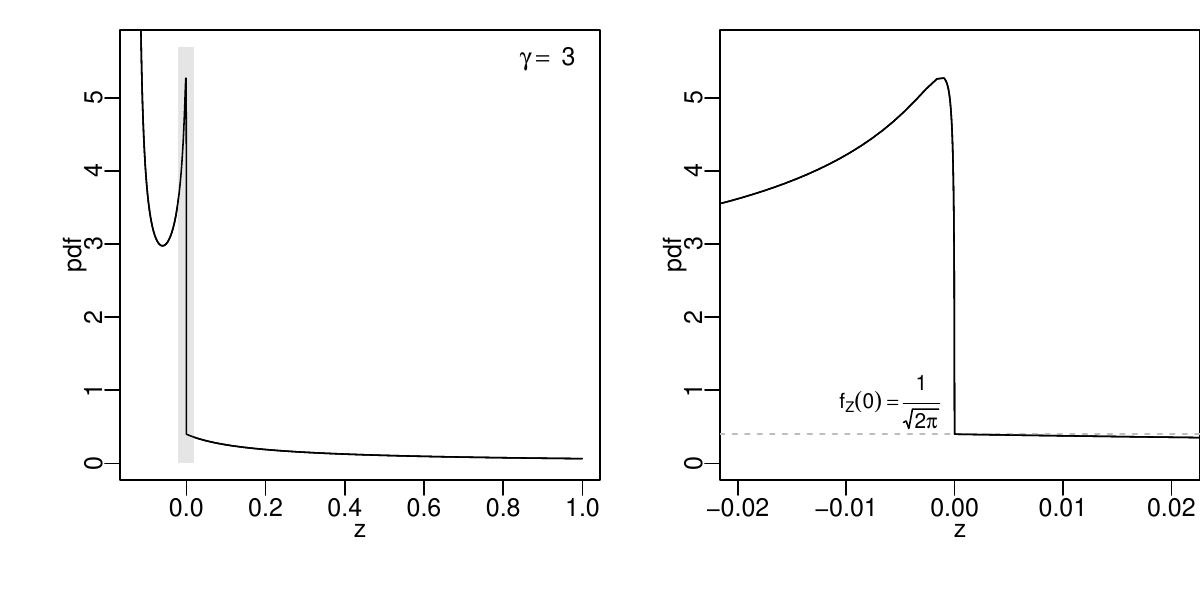}
\caption{Example of Lambert $W\times N(0, 1)$ pdf with $\gamma = 3$. The~right panel shows a closer view of the interval marked with grey in the left~panel.}
\label{fig:WNorm-gamma3}
\end{figure}

The point $-\frac{1}{\gamma e}$ where $f_Z$ has an asymptote can be thought of as a point where the transformation folds the left tail of $N(0, 1)$ and  fits it into the interval $(-\frac{1}{\gamma e}, 0)$.
At this turning point, the~density {accumulates}; see Figures~\ref{fig:WNorm-pdf-cdf}--\ref{fig:WNorm-gamma25} for~examples.
Although the transformation squeezes the negative values of $N(0, 1)$ into a fixed interval and makes the right tail heavier, it continues to have zero as a point where the probability mass is divided into equal halves. Furthermore, at point $z=0$, the~pdf $f_Z$ is equal to the pdf of $N(0, 1)$, i.e.,~$f_Z(0)=\frac{1}{\sqrt{2\pi}}$. This property is pointed out in the right-hand panels of Figures~\ref{fig:WNorm-gamma25} and \ref{fig:WNorm-gamma3}.

\unskip

\begin{Lemma}
The principal branch component of the pdf of a Lambert $W\times N(0,1)$ random variable $f_0$ has the following properties. The~function $f_0(z)$:
\begin{itemize}
\item[(a)] has two local extrema (maximum and minimum) if $\gamma\in (0,\sqrt{2}-1)$; and
\item[(b)] is monotone decreasing if $\gamma > \sqrt{2}-1$.
\end{itemize}
\end{Lemma}

\begin{Lemma}
The non-principal branch component of the pdf of a Lambert $W\times N(0,1)$ random variable $f_{-1}$ has the following properties. The~function $f_{-1}(z)$:
\begin{itemize}
\item[(a)]  is monotone increasing (to 0) if $\gamma\in (0,\sqrt{2}+1)$; and
\item[(b)]  has two local extrema (maximum and minimum) if $\gamma > \sqrt{2}+1$.
\end{itemize}
\end{Lemma}

Consequently, depending on the value of the skewness parameter $\gamma$, it is possible to distinguish three main shapes of the pdf of a Lambert $W$ normal random variable. 
First, if \mbox{$\gamma \in (0, \sqrt{2}-1)$,} then the~pdf has two local extrema due to the principal branch component $f_0$; see Figure~\ref{fig:WNorm-gamma0203} or the left panel of Figure~\ref{fig:WNorm-gamma0405} for~examples. 
Second, if~$\gamma \in [\sqrt{2}-1,\sqrt{2}+1]$, then the~pdf is {a} strictly decreasing function of $z$, as~in the right panel of Figure~\ref{fig:WNorm-gamma0405}.
Third, if~$\gamma > \sqrt{2}+1$, then the~pdf again has two local extrema, now due to the non-principal branch component $f_{-1}$, and~compared to the first case, the~overall shape of pdf is different, as seen in Figures~\ref{fig:WNorm-gamma25} and~\ref{fig:WNorm-gamma3}. In~these two figures, the~right panel provides a more detailed view of the interval where the maximum is placed. {As notably seen in Figure} \ref{fig:WNorm-gamma3}, the~apparently sharp peak turns out to be quite smooth if examined more~closely.

\begin{figure}[H]
\includegraphics[width=0.81\textwidth]{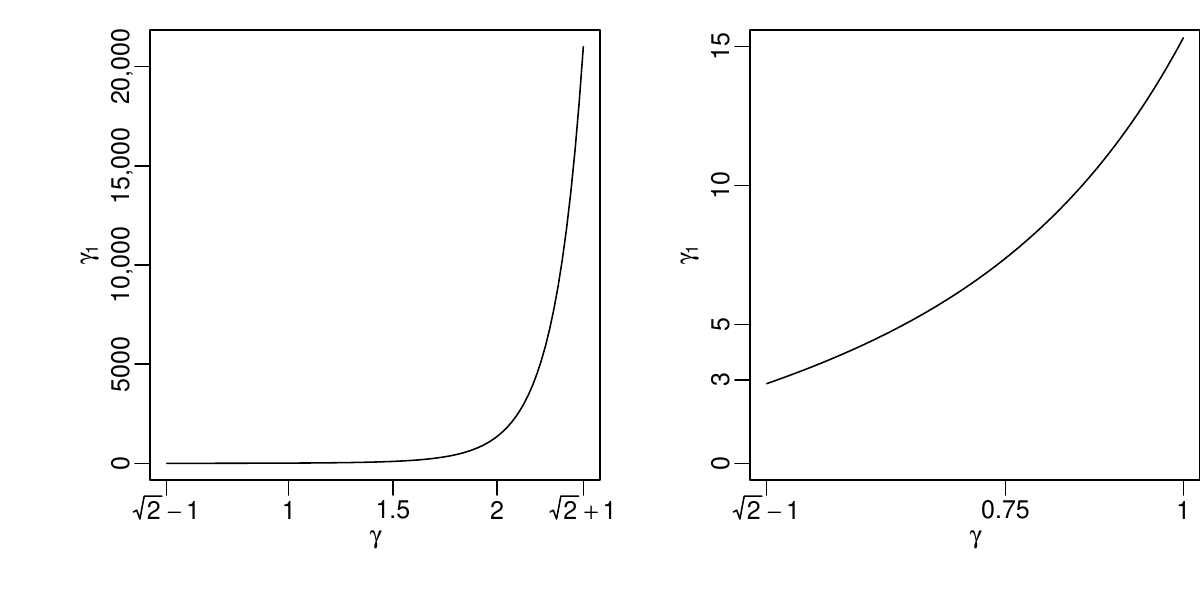}
\caption{{Skewness} 
 coefficient $\gamma_1$ for Lambert $W\times N(\mu, \sigma)$ random variables for different ranges of the skewness parameter $\gamma$.}
\label{fig:WNorm-skew-coef}
\end{figure}

Lastly, we provide the expressions of the moments and skewness coefficient of a Lambert $W\times N(\mu, \sigma)$ random variable.
The moments of a Lambert $W\times N(0,1)$ random variable can be found using the moment generating function (mgf) of the underlying standard normal distribution.
Let $Z$ be a Lambert $W\times N(0,1)$ random variable. {The moments for $Z$ are then as follows }\citep{Goerg2011}:

 \begin{equation*}
 E(Z^k)=\frac{1}{k^{k}}\frac{\partial^{k}}{\partial\gamma^{k}} M_{N(0,1)}(\gamma k)=
 \frac{1}{k^{k}}\frac{\partial^{k}} {\partial \gamma^{k}}\exp\left({\frac{\gamma^{2}k^{2}}{2}}\right),
 \end{equation*}
 where $M_{N(0, 1)}$ denotes the mgf of $N(0, 1)$. For~the general case, i.e.,~for a Lambert $W\times N(\mu, \sigma)$ random variable $Y$, we can use the properties of the location-scale family, meaning that we have  
 \begin{equation*}
     E(Y^k) = E((Z\sigma+\mu)^k).
 \end{equation*}
{As the} 
 moments are found using the derivatives of an exponential function, the~moments of any order $k$ exist and are finite.
Using the above expressions, we can derive the formulae for the mean of $Y$
\begin{equation}
\label{eq:WnormEY}
EY = \mu+\sigma \gamma e^{\gamma^2/2},
\end{equation}
the variance of $Y$
\begin{equation}
\label{eq:WnormDY}
{Var}Y=\sigma^2 e^{\gamma^2}(e^{\gamma^2}(1+4\gamma^2)-\gamma^2),
\end{equation}
and  the skewness coefficient $\gamma_{1}(Y)$: 
\begin{equation}\label{eq:Wnorm-skewness}
\gamma_{1}(Y)= \gamma\left(\frac{e^{3\gamma^{2}} (9+27\gamma^{2})-e^{\gamma^{2}}(3+12\gamma^{2})+2\gamma^{2}} {(e^{\gamma^{2}}(1+4\gamma^{2})-\gamma^{2})^{\frac{3}{2}}}\right).
\end{equation}
{The} 
 skewness coefficient is a monotone function of $\gamma$, and has~the same sign. As~$\gamma \rightarrow \pm \infty$, we have $\gamma_1(Y) \rightarrow \pm \infty$, and~the speed of growth is exponential. For~example, if~we look at the range of values $\gamma\in(\sqrt{2}-1; \sqrt{2}+1)$, where the pdf is monotone decreasing, the~skewness coefficient grows from around 3 to 20,000 (see Figure~\ref{fig:WNorm-skew-coef}). 

\unskip
\unskip
\unskip

\subsection{Lambert $W$ Exponential~Distribution}
Let $X$ be an exponentially distributed random variable with parameter $\lambda>0$ ($\lambda$ as rate). Then, the transformed random variable
\[
Y = X e^{\gamma\lambda X}
\]
has a Lambert $W\times Exp(\lambda)$ distribution with parameter vector $(\lambda,\gamma)$. According to \eqref{eq:skaala}, for~positive $\gamma$, the~cdf of $Y$ is
\begin{equation*}
F_{Y}(y\vert \lambda,\gamma)=
1-\exp{\left(- \frac{W_{0}(\gamma \lambda y)}{\gamma}\right)}, \quad  y\geq0,
\end{equation*}
and, using \eqref{eq:scale-pdf}, the~pdf of $Y$ is
\begin{equation*}
f_{Y}(y\vert \lambda,\gamma)=
\lambda\exp{\left(- \frac{W_{0}(\gamma \lambda y)}{\gamma}\right)} \frac{\exp{(-W_{0}(\gamma \lambda y))}}{1+W_{0}(\gamma \lambda y)}, \quad y\geq0.
\end{equation*}


For $\gamma <0$, the~expressions for cdf and pdf additionally involve the non-principal branch of the Lambert $W$ function, as seen in \eqref{eq:scale-cdf2} and \eqref{eq:scale-pdf2}:
\begin{equation*}
F_{Y}(y\vert \lambda,\gamma)=
1-\exp{\left(- \frac{W_{0}(\gamma \lambda y)}{\gamma}\right)} + \exp{\left(- \frac{W_{-1}(\gamma \lambda y)}{\gamma}\right)}  , \quad   0\leq y < -\frac{1}{e\gamma\lambda},
\end{equation*}
and
\begin{multline*}
f_{Y}(y\vert \lambda,\gamma)=
\lambda \exp{\left(- \frac{W_{0}(\gamma \lambda y)}{\gamma}\right)} \frac{\exp{(-W_{0}(\gamma \lambda y))}}{1+W_{0}(\gamma \lambda y)}\\- 
\lambda \exp{\left(- \frac{W_{-1}(\gamma \lambda y)}{\gamma}\right)} \frac{\exp{(-W_{-1}(\gamma \lambda y))}}{1+W_{-1}(\gamma \lambda y)}{,}
\quad  0\leq y < -\frac{1}{e\gamma\lambda}.
\end{multline*}

\begin{figure}[H]
\includegraphics[width=0.93\textwidth]{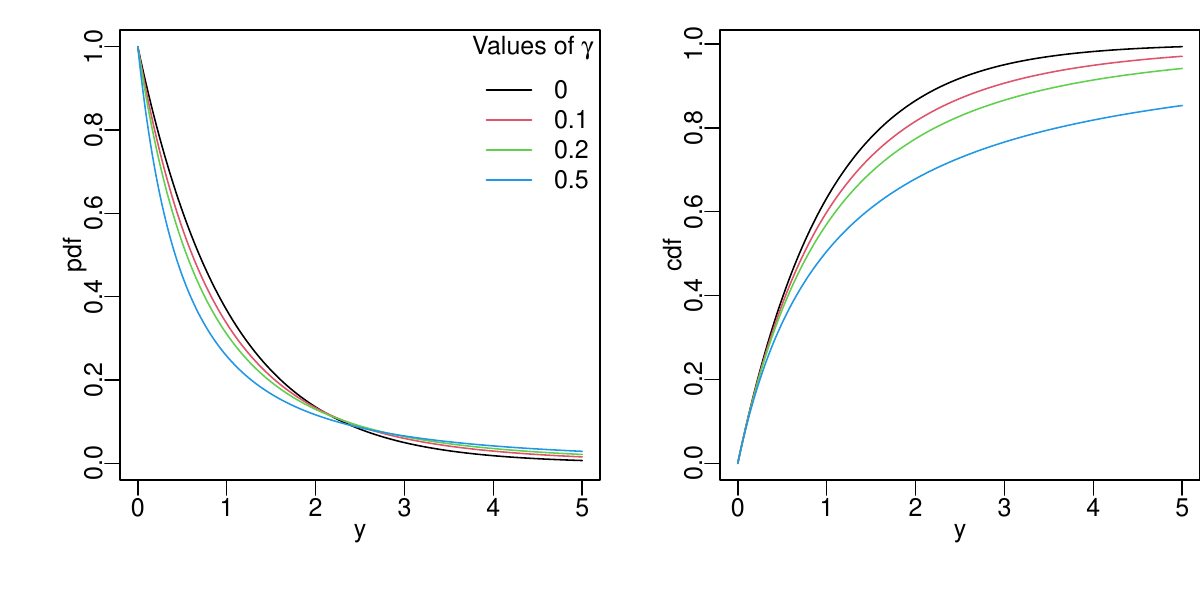}
\caption{Plots of the pdf (left panel) and cdf (right panel) of $W\times Exp(1)$ distributions with different positive $\gamma$ values\label{fig:WExp-pos-gamma}.}
\end{figure}

\begin{figure}[H]
\includegraphics[width=0.93\textwidth]{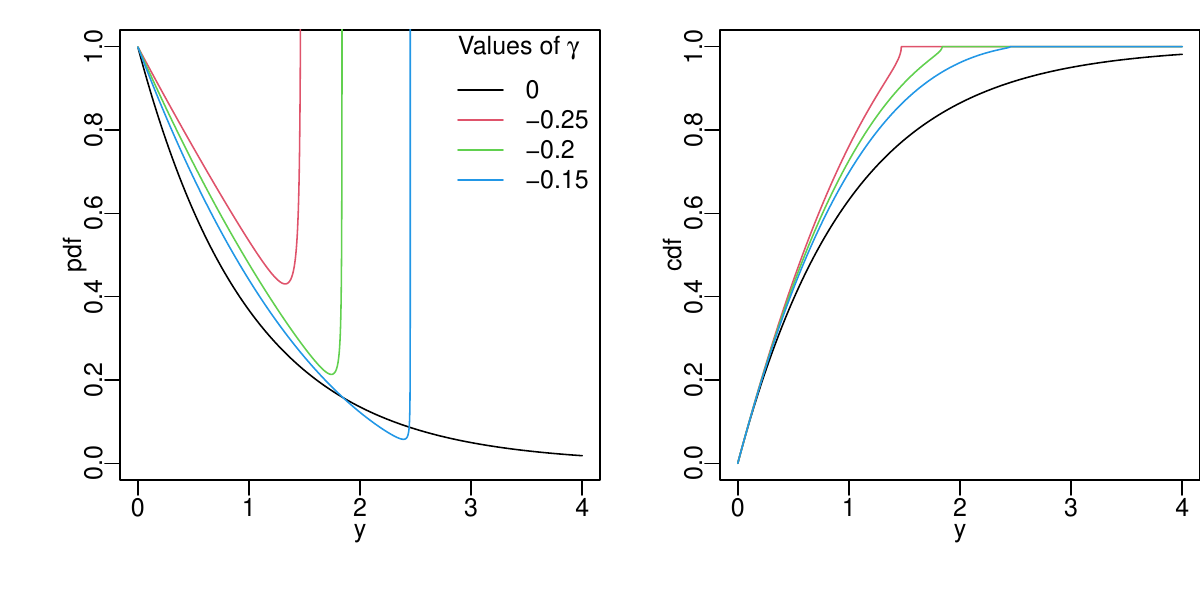}
\caption{Plots of the pdf (left panel) and cdf (right panel) of $W\times Exp(1)$ distributions with different negative $\gamma$ values\label{fig:WExp-neg-gamma}.}
\end{figure}

{For} 
 examples of the pdf and cdf for the Lambert $W\times Exp(1)$ distribution, see Figures~\ref{fig:WExp-pos-gamma} and~\ref{fig:WExp-neg-gamma}. As~apparent from Figure~\ref{fig:WExp-pos-gamma}, the Lambert random variables have a heavier tail in the case of positive $\gamma$ as compared to the exponential distribution.


For negative $\gamma$ values (see Figure~\ref{fig:WExp-neg-gamma}), the~random variable $Y$ takes values in the fixed interval $(0, -\frac{1}{e\gamma\lambda})$, as the transformation relocates the larger values of the underlying exponential random variable $X$.
While it can be argued that this kind of transformation is not relevant for typically heavy-tailed insurance data, our example (see 
Section~\ref{sec:LambertWapp})  
shows an adequate fit when using the Lambert $W$ exponential random variables with $\gamma<0$ for log claims of Danish fire loss data. In~the case of $\gamma <0$, if~the absolute value of $\gamma$ is small, this produces a distribution with a suitably large cut-off point to fit data with moderate tails, as is the case for the Danish log claims data. Similarly, only small values of $\gamma$ are of practical use for~positive $\gamma$, as~the tail quickly becomes heavy very. For~example, if~$\gamma \geq 1$, then Lambert  $W\times Exp(\lambda)$ random variables do not have a finite first moment. For~$\gamma <1$, the~first moment is $\frac{1}{\lambda(1-\gamma)^2}$. In~general, the~following expression holds:
\begin{equation*}\label{eq:WExp-momendid}
EY^k = \frac{k!}{\lambda^k(1-k\gamma)^{k+1}},\quad \textup{ if } \quad\gamma < \frac1k.
\end{equation*}

{The} 
 skewness coefficient for a Lambert $W\times Exp(\lambda)$ random variable with  $\gamma<\frac13$ can be calculated~as follows:
\begin{multline}
\label{eq:WxEgamma1}
\gamma_1(Y) = 2\sqrt{\frac{(1-2\gamma)^9}{(2\gamma^4-2\gamma+1)^3}}
\left(\frac{3(1-\gamma)^4((1-\gamma)^2(1-2\gamma)^3 - (1-3\gamma)^4)}{(1-3\gamma)^4(1-2\gamma)^3}+1\right).
\end{multline}

\begin{figure}[H]
\includegraphics[width=0.6\textwidth]{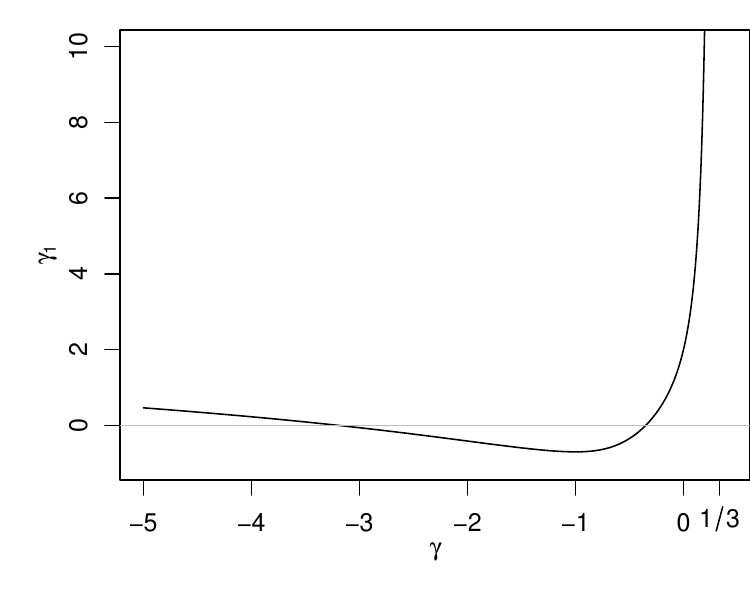}
\caption{Skewness coefficient $\gamma_1$ for Lambert $W\times Exp(\lambda)$ with parameter $\gamma \in (-5, \frac13)$.\label{fig:WExp-skewness}}
\end{figure}

{If} 
 $\gamma \geq \frac13$, then the~third moment of $Y$ is infinite, and~the coefficient $\gamma_1(Y)$ cannot be found. 
The skewness coefficient is a non-monotonic function of $\gamma$ (see Figure~\ref{fig:WExp-skewness}).
If $\gamma=0$, the~distribution simplifies to exponential, and~the skewness coefficient  $\gamma_1 = 2$. For~a Lambert $W\times Exp(\lambda)$ distribution, the~skewness coefficient $\gamma_1$ can exceed {a value of 2}, and approaches infinity as $\gamma \rightarrow \frac13$. For~values $-\infty<\gamma<-1$, the~skewness coefficient is a decreasing function of $\gamma$ with a minimum value of $-\frac{9\sqrt{15}}{50}$, while~for $\gamma\in(-1,\frac13)$ it is increasing (see Figure~\ref{fig:WExp-skewness}).

\unskip
\unskip
\unskip


\section{Fitting Lambert $W$ Random Variables to Insurance~Data}\label{sec:LambertWapp}
In this subsection, we fit the Lambert $W$ normal and exponential random variables on two well-known datasets, the~US indemnity data introduced in \citep{frees1998} and the Danish fire data introduced in \citep{mcneil1997}, then~compare the fit with previous~results.

These datasets have been widely used in field-specific literature before; see, \mbox{e.g.,~\citep{Klugman1999,Dupuis2006}} for the US indemnity data and~\citep{Resnick1997,Cooray2005, DellAquila2006} for the Danish fire data, among~others. A~consolidated overview of previous results is provided in \citep{eling2012}.

\begin{figure}[H]
\includegraphics[width=0.8\textwidth]{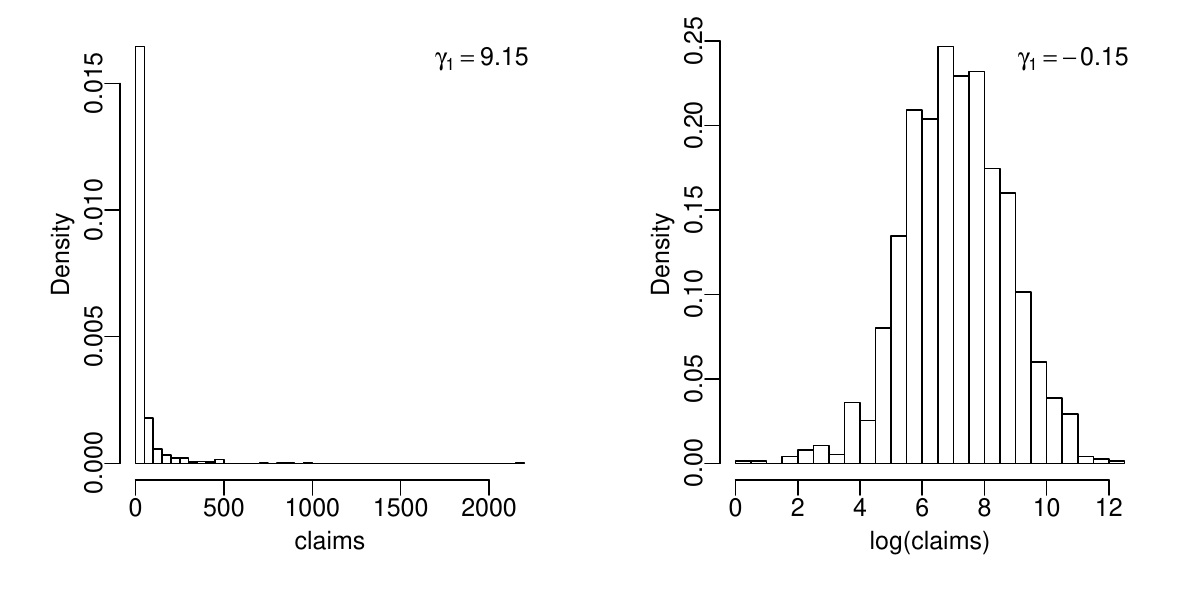}
\caption{Left panel: US indemnity data (in thousands of USD). Right panel: same data \linebreak after~log-transformation.}
\label{fig:data-usa}
\end{figure}

\begin{figure}[H]
\includegraphics[width=0.8\textwidth]{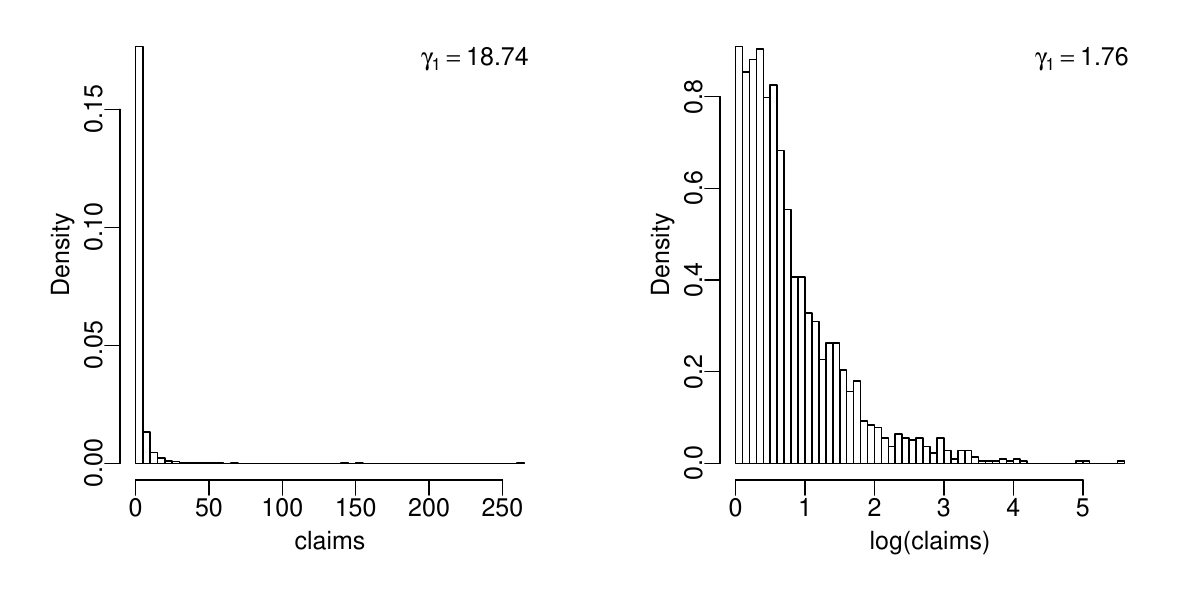}
\caption{Left panel: Danish fire data claims (in millions of DKK). Right panel: same data \linebreak after~log-transformation.}
\label{fig:data-taani}
\end{figure}

To recall the distributions of these example datasets, see Figure~\ref{fig:data-usa} for the US indemnity data and Figure~\ref{fig:data-taani} for the Danish fire loss data. 
In both figures, the~left panel presents the data on the original scale (thousands of USD for US indemnity and millions of DKK for Danish fire data), and the right panel presents the same data after log transformation.  In~the case of the log-transformed data, we use a similar shift to the one in \citep{eling2012} in order to keep the results comparable. More precisely, the~transformation $\ln(y) - \min(\ln(y)) + 10^{-10}$ is applied  on the original variable~$y$.

{It is evident that both datasets exhibit significant skewness when observed} on the original scale. The~skewness is more extreme for~the Danish fire data, with a skewness coefficient of $\gamma_1=18.74$, as compared to  $\gamma_1=9.15$ for the US indemnity data. In~the case of the US indemnity data, the~log-transformed data produces an almost symmetric histogram that is very similar to a normal distribution. The~log-transform reduces the skewness for Danish fire data as well, although~the result remains skewed, with~$\gamma_1=1.76$.

\unskip

In \citep{eling2012}, nineteen distributions were fitted to the two aforementioned datasets, with~the result that the skew-normal and skew $t$ distributions are reasonably competitive compared to other models commonly used for insurance~data.

In our research, we follow this construction and include all fitted continuous distributions while adding three more distributions to the list: the Lambert $W$ normal and exponential distributions as our main contribution, and~the Pareto distribution, which was previously missing due to technical problems. We use the maximum likelihood method for parameter estimation, as in \citep{eling2012}. For~more details of the estimation process, see Appendix \ref{sec:app_estim}.

To compare these models with competitors, we measure the goodness of fit between the data and distribution using the Akaike Information Criterion (AIC) and the Bayesian Information Criterion (BIC). The BIC is included because the number of parameters of the distributions ranges from 1 to 5, making the penalty of the AIC quite small compared to the flexibility that additional parameters can~provide.

Before the comparison, we first examine the parameter estimates of the Lambert $W$ distributions in Table~\ref{tab:par-hinnangud}.

\begin{table}[H]
\caption{{Parameter} 
 estimates for Lambert~distributions.\label{tab:par-hinnangud}}
\newcolumntype{C}{>{\centering\arraybackslash}X}
\begin{tabularx}{\textwidth}{p{3cm}CC|CCC}
\toprule
 \multirow{2}{*}{\textbf{Data}\vspace{-6pt}}  & \multicolumn{2}{c}{\boldmath{$W \times Exp(\lambda)$}} & \multicolumn{3}{c}{\boldmath{$W \times N(\mu, \sigma)$}} \\
\cmidrule{2-6}
  & \boldmath{$\lambda$} & \boldmath{$\gamma$} & \boldmath{$\mu$} &\boldmath{$\sigma$} & \boldmath{$\gamma$}\\
\midrule
US indemnity & 0.080 & \phantom{$-$}0.496 & 13.444 & 28.829 & \phantom{$-$}0.789\\
US indemnity, log & 0.093 & $-$0.321 & \phantom{1}7.106 & \phantom{2}1.635 & $-$0.021\\
\midrule
Danish fire & 0.386 & \phantom{$-$}0.096 & \phantom{1}1.923 & \phantom{2}1.417 & \phantom{$-$}0.564\\
Danish fire, log & 1.176 & $-$0.040 & \phantom{1}0.542 & \phantom{2}0.549 & \phantom{$-$}0.373\\
\bottomrule
\end{tabularx}
\end{table}
In the case of the Lambert $W$ exponential model for the US indemnity data, the $\gamma$ estimate $0.496$ provides an infinite skewness coefficient. At~the same time, the~fit according to the BIC is good relative to other models; see {Table}~\ref{tab:usa}  
and later discussion. As~the US indemnity data are close to normal on the log scale, the~Lambert $W$ exponential is not really a suitable model here. However,~the estimate $\hat\gamma=-0.321$ {corresponds} to a skewness coefficient value of $0.09${, i.e.} this model is able to pick up the symmetry of the data.

What is interesting in the case of the Danish log data is the negative $\gamma$ estimate, as~{it} produces a distribution with an upper bound of $-\frac{1}{\gamma e \lambda}$, here resulting in $7.82$. As~the maximum value in data is around $5.57$, this model allows {even} higher claim values {than} in the data. Furthermore, this model suits the data well, as~it ranks high according to BIC value (see Table~\ref{tab:taani}, discussed in detail later on). The~fit is not good for~the same data on the original scale, which are highly skewed, and the $\gamma$ estimate $0.096$ can be considered unexpectedly low.

From the estimates of the Lambert $W$ normal parameter, we can point out that the~$\gamma$ estimates are in the interval that produces monotone decreasing pdf for both datasets on the original scale. For~US indemnity data on the log scale, the~$\gamma$ estimate $-0.021$ produces a distribution very similar to normal{,} which is in agreement with the histogram. For~the Danish log data, the~estimate for $\gamma$ is $0.373$, {which} is in the interval $(0, \sqrt{2} - 1)$, which corresponds to the pdf shape with some downward bend between the asymptote and maximum, as in the left panel of Figure~\ref{fig:WNorm-gamma0405}. As~shown in the following analysis, the~fit provided by the Lambert $W$ transformed random variables is~promising.

The results of model fitting are presented in Tables~\ref{tab:usa} and \ref{tab:taani}. The~distributions are sorted in ascending order by the number of parameters, with~the two newly added Lambert $W$ distributions always shown at the top of the table. In~every column, the~first three results are marked: the best result is in bold, the~second-best is underlined, and~the third-best is underlined and in~italics. 

\begin{table}[H]
\caption{{US indemnity data:} 
AIC and BIC values for fitted~distributions.}\label{tab:usa}
\newcolumntype{C}{>{\centering\arraybackslash}X}
\begin{tabularx}{\textwidth}{p{3.5cm}cCCCC}
\toprule
\multirow{2}{*}{\textbf{Distribution}\vspace{-6pt}} & & \multicolumn{2}{c}{\textbf{AIC}} & \multicolumn{2}{c}{\textbf{BIC}} \\
\cmidrule{2-6}
  & \textbf{Npar} & \textbf{Original} & \textbf{Log}& \textbf{Original}& \textbf{Log}\\
\midrule
Lambert $W$ exponential & 2 & \underline{13141.92} & 7845.81 & \underline{13152.55} & 7856.44\\
Lambert $W$ normal & 3 & 13397.48 & \underline{5737.79} & 13413.42 & \underline{\em{5753.73}}\\
\midrule
exponential & 1 & 14157.93 & 8869.95 & 14163.24 & 8875.26\\
gamma & 2 & 13537.17 & 6442.22 & 13547.80 & 6452.85\\
log-normal & 2 & \textbf{13137.53} & 8895.12 & \textbf{13148.16} & 8905.74\\
logistic & 2 & 16544.91 & 5753.92 & 16555.54 & 5764.55\\
normal & 2 & 18156.65 & 5740.44 & 18167.27 & \textbf{5751.06}\\
Weibull & 2 & 13321.70 & 5923.95 & 13332.33 & 5934.58\\
Cauchy & 2 & 14518.07 & 6264.44 & 14528.69 & 6275.07\\
Pareto & 2 &  13148.51 & 8871.95 &    13159.13 &  8882.58\\
symm hyperbolic & 3 & 15884.38 & 5738.41 & 15900.32 & 5754.35\\
symm NIG \textsuperscript{1} & 3 & 14515.76 & 5738.38 & 14531.70 & 5754.32\\
symm VG \textsuperscript{2} & 3 & 14261.53 & 5738.65 & 14277.47 & 5754.59\\
student $t$ & 3 & 14492.64 & 5738.12 & 14508.58 & 5754.06\\
skew-normal & 3 & 16315.13 & \textbf{5737.79} & 16331.07 & \underline{5753.73}\\
asymm hyperbolic & 4 & 14163.24 & 5738.16 & 14184.49 & 5759.41\\
asymm NIG & 4 & 13148.66 & 5738.12 & \underline{\em{13169.91}} & 5759.37\\
asymm VG & 4 & 14177.46 & 5738.61 & 14198.71 & 5759.86\\
symm ghyp \textsuperscript{3} & 4 & 14494.64 & 5740.43 & 14515.89 & 5761.68\\
skew $t$ & 4 & 13197.79 & \underline{\em{5738.06}} & 13219.05 & 5759.32\\
asymm ghyp & 5 & \underline{\em{13145.91}} & 5740.61 & 13172.48 & 5767.17\\
\bottomrule
\end{tabularx}
{\footnotesize{\textsuperscript{1} normal inverse Gaussian; \textsuperscript{2} variance gamma; \textsuperscript{3} generalized hyperbolic.}}
\end{table}

In the case of the US indemnity data (see Table~\ref{tab:usa}), we have seen earlier that the log-transformed data closely resemble the normal distribution. Therefore, the log-normal distribution can be expected to provide the best fit for the data in the original scale. However,~the Lambert $W$ exponential model provides a good fit as well, with the second-best AIC and BIC values. For~the log-transformed data, the~two smallest AIC values are almost equal, with~the following block having very close values. Thus, the skew-normal and Lambert $W$ normal distributions share first place, and skew $t$ follows at the top of the next block. Based on the BIC, the normal distribution provides the best fit, having fewer parameters than the skew-normal or Lambert $W$ normal. The skew-normal and Lambert $W$ normal distributions fall to second and third place, respectively. The~pdfs for the best three models with data histograms are plotted in Figure~\ref{fig:data-usa-3joont} in Appendix \ref{sec:app_joonised}. {It is apparent from the latter graph that the top three models exhibit a high degree of similarity, with~the primary distinction residing in the region of small claims when viewed on the original scale.} For the log-transformed data, the~three curves practically~coincide.
 
\begin{table}[H] 
\caption{{Danish fire data:} AIC and BIC values for fitted~distributions.}\label{tab:taani}
\newcolumntype{C}{>{\centering\arraybackslash}X}
\begin{tabularx}{\textwidth}{p{3.5cm}cCCCC}
\toprule
\multirow{2}{*}{\textbf{Distribution}\vspace{-6pt}} & & \multicolumn{2}{c}{\textbf{AIC}} & \multicolumn{2}{c}{\textbf{BIC}} \\
\cmidrule{2-6}
  & \textbf{Npar} & \textbf{Original} & \textbf{Log}& \textbf{Original}& \textbf{Log}\\
\midrule
Lambert $W$ exponential & 2 & 9264.10 & 3282.22 & 9275.46 & \underline{3293.58}\\
Lambert $W$ normal  & 3 & \underline{6699.82} & \textbf{2978.46} & \underline{6716.86} & \textbf{2995.50}\\
\midrule
exponential & 1 & 9620.79 & 3297.61 & 9626.47 & 3303.30\\
gamma & 2 & 9538.19 & 3299.61 & 9549.55 & 3310.98\\
log-normal & 2 & 8119.79 & 5504.62 & 8131.16 & 5515.98\\
logistic & 2 & 11479.71 & 4421.17 & 11491.08 & 4432.53\\
normal & 2 & 15431.52 & 4709.15 & 15442.89 & 4720.52\\
Weibull & 2 & 9611.24 & 3294.27 & 9622.61 & 3305.63\\
Cauchy & 2 & 8240.17 & 4589.38 & 8251.53 & 4600.74\\
Pareto & 2&    9249.67 &   3818.07   &9261.03&3829.43 \\
symm hyperbolic & 3 & 10433.17 & 4363.90 & 10450.21 & 4380.95\\
symm NIG \textsuperscript{1}& 3 & 8237.61 & 4303.93 & 8254.66 & 4320.97\\
symm VG \textsuperscript{2} & 3 & 9089.69 & 4375.17 & 9106.73 & 4392.21\\
student $t$ & 3 & 8237.85 & 4299.90 & 8254.90 & 4316.94\\
skew-normal & 3 & 12608.36 & 3441.49 & 12625.40 & 3458.54\\
asymm hyperbolic & 4 & 8109.27 & 3307.83 & 8132.00 & 3330.56\\
asymm NIG & 4 & 6806.79 & 3378.14 & 6829.52 & 3400.86\\
asymm VG & 4 & 7404.07 & \underline{\em{3281.06}} & 7426.80 & 3303.78\\
symm ghyp \textsuperscript{3} & 4 & 8224.65 & 4298.21 & 8247.38 & 4320.93\\
skew $t$ & 4 & \textbf{6683.02} & \underline{3274.24} & \textbf{6705.75} & \underline{\em{3296.96}}\\
asymm ghyp & 5 & \underline{\em{6775.85}} & 3283.06 & \underline{\em{6804.26}} & 3311.46\\
\bottomrule
\end{tabularx}
{\footnotesize{\textsuperscript{1} normal inverse Gaussian; \textsuperscript{2} variance gamma; \textsuperscript{3} generalized hyperbolic.}}
\end{table}

From Table~\ref{tab:taani}{,} it can be seen that for the Danish fire data on the original scale, the~two best-fitting models are the skew $t$ and Lambert $W$ normal distributions.
For the Danish log data, the~Lambert $W$ normal distribution again has the best fit based on the AIC, followed by the skew $t$. Based on the BIC, the~best model is the Lambert $W$ normal distribution, while the~Lambert $W$ exponential has the second best result; for~further illustration{,} see Figure~\ref{fig:data-taani-3joont} in Appendix \ref{sec:app_joonised}. The three best pdfs for the original data are very similar. On~the log-transformed data, the~discrepancies are not large either, though~{they are} more clearly visible. In~conclusion, the~Lambert $W$ models provide a good fit to both the original and log-transformed~data.

\section{Summary}

In this paper, we have addressed the Lambert $W$ transform-based approach and the properties of the resulting distributions, thoroughly investigating the Lambert $W$ normal and Lambert $W$ exponential distributions. We introduce the skewness via the Lambert $W$ transform and the skewness parameter $\gamma$. Without loss of generality, we focus on positive values of $\gamma$, as these are more of interest in loss modelling applications. For~the Lambert $W$ standard normal distribution with a positive skewness parameter $\gamma$, the~pdf $f(y)$ has an asymptote at $y=-\frac{1}{\gamma e}$. We establish the following three regions based on the shape of the~pdf:
\begin{itemize}
    \item[(a)] If $\gamma \in (0, \sqrt{2} - 1)$, then the~pdf has two local extrema;
    \item[(b)] If $\gamma \in (\sqrt{2} - 1, \sqrt{2} + 1)$, then the~pdf is monotone decreasing;
    \item [(c)] If $\gamma > \sqrt{2} {+} 1$, then the~pdf has two local extrema.
\end{itemize}

In the first range, where $\gamma \in (0, \sqrt{2} - 1)$, the~shape of the distribution is at~first glance not the most suitable for loss modelling, {and} needs additional explanations.
Nevertheless, it can be argued that the asymptote effect is reasonably small, meaning that the distribution can provide a good fit, as in~the Danish fire log data. Such a shape might be suitable in zero-altered models as well, where zero claims are included.
The second and most appealing range, where the pdf is monotone decreasing, covers a wide range of the skewness coefficient values; see Figure~\ref{fig:WNorm-skew-coef}. If~$\gamma = \sqrt{2} + 1$, then the~skewness coefficient is about 20,000; thus, the not-very-suitable shape in the third range is not a problem for most practical applications. 

For the Lambert $W$ exponential distribution, we establish that it allows a wider choice of the skewness coefficient than the exponential distribution. Moreover, one additional parameter relaxes the rigid relationship between the mean and variance of the exponential distribution. These properties make the Lambert exponential distribution a promising model for insurance loss~data.

Our results in the practical part show that the Lambert $W$ transformed distributions operating in a wide range of skewness represent a viable choice for insurance loss modelling. Both the normal and exponential distribution-based transforms show a reasonably good fit. An~especially illustrative proof of this flexibility is visible in the Danish fire data, where the results of the Lambert $W$ normal model are well at the top for both the original and log-transformed datasets.

Clearly, the~choices available for the Lambert $W$ approach are not limited to normal and exponential random variables. While the normal and exponential distributions seem to be a natural starting point for loss modelling, other distributions can offer \mbox{valuable~contributions as well.}

\vspace{6pt}

\authorcontributions{Conceptualization, M.K. and A.S.; 
methodology, M.K.,  A.S., and T.P.; 
software, A.S., T.P., and L.T.; 
validation, A.S.; 
formal analysis, all authors; 
data curation, A.S.; 
writing---original draft preparation, M.K., T.P., and L.T.; 
writing---review and editing, M.K. and A.S.; visualization, A.S. and T.P.; 
supervision, M.K.; 
project administration, A.S. and M.K.; 
funding acquisition, M.K.
{All authors have read and agreed to the published version of the manuscript.}}

\funding{{This work} 
 was supported by the Estonian Research Council, grant~PRG1197.}

\dataavailability{We used the R package \emph{fExtremes} \citep{pakett-fExtremes} to access the US indemnity data and the package \emph{copula} \citep{pakett-copula} for the Danish fire loss data.} 

\acknowledgments{The authors are thankful to Roel Verbelen for constructive discussions and comments on an earlier draft of the paper. The authors also thank all the anonymous referees for their valuable and constructive~feedback.}

\conflictsofinterest{The authors declare no conflicts of~interest.}

\appendixtitles{yes} 
\appendixstart
\appendix

\section[\appendixname~\thesection]{Proofs of the Properties of Lambert $W$ Standard Normal Random Variables}
\label{sec:app_proofs}

In this appendix, we provide the proofs of the properties of the Lambert $W$ standard normal random variables formulated in Lemmas 1--3 in Section~\ref{sec:LambertWN}.

\begin{proof}[Proof of Lemma 1]
Recall that the density $f_Z(z)$ can be expressed as $f_0(z)-f_{-1}(z)$ for $z\in (-\frac{1}{\gamma e},0]$, where $f_0$ and $f_{-1}$ are the principal and non-principal branch components of the pdf, respectively.

Furthermore, recall the form of the principal branch component $f_0(z)$ as specified in \eqref{eq:f0}:
\begin{equation*}
f_0(z) = \frac{1}{\sqrt{2\pi}}\exp\left({-\frac{(W_{0}(\gamma z))^2}{2\gamma^2}}\right)\frac{\exp({-W_{0}(\gamma z)})} {1+W_{0}(\gamma z)}
\end{equation*}
for $z>-\frac{1}{\gamma e}$.

Looking separately at the components of this expression, it is easy to see that \linebreak$W_{0}(\gamma z)\to -1$, $(W_{0}(\gamma z))^{2}\to 1$, and~$1+W_{0}(\gamma z)\to 0+$ if~$z\to -\frac{1}{\gamma e}+$. Thus, the~ratio $\displaystyle\frac{\exp({-W_{0}(\gamma z)})} {1+W_{0}(\gamma z)}$ tends to infinity in the process, which implies that the principal branch component $f_0(z)$ specified in  \eqref{eq:f0} goes to infinity if $z\to-\frac{1}{\gamma e}+$.

A similar argument holds for the non-principal branch component $f_{-1}(z)$.
First, recall that, as~stated in Formula  \eqref{eq:fm1}, the~non-principal branch component has the following~form:
\begin{equation*}
f_{-1}(z) = \frac{1}{\sqrt{2\pi}}\exp\left({-\frac{(W_{-1}(\gamma z))^2}{2\gamma^2}}\right)\frac{\exp({-W_{-1}(\gamma z)})} {1+W_{-1}(\gamma z)}
\end{equation*}
with $z\in (-\frac{1}{\gamma e},0]$.

Analyzing the components of this expression separately, it can be seen that \linebreak $W_{-1}(\gamma z)\to -1$, $(W_{-1}(\gamma z))^{2}\to 1$, and~$1+W_{-1}(\gamma z)\to 0-$ in the process where $z\to-\frac{1}{\gamma e}+$. This implies that $\displaystyle \frac{\exp({-W_{-1}(\gamma z)})} {1+W_{-1}(\gamma z)}\to -\infty$, which, in~summary, results in $\lim_{z\to-\frac{1}{\gamma e}+}f_{-1}(z) = -\infty$.

In conclusion, because $f_Z(z) = f_0(z)-f_{-1}(z)$  for $z\in (-\frac{1}{\gamma e},0]$, we have\linebreak $\lim_{z\to-\frac{1}{\gamma e}+}f_Z(z) = \infty$. The~lemma is~proved.
\end{proof}

\begin{proof}[Proof of Lemma 2]    
We first note that Formulas~\eqref{eq:f0} and \eqref{eq:fm1} differ only in the specification of the branch ($W_0$ or $W_{-1}$). Because most of the following argumentation holds for both branches, we do not specify the branch unless explicitly needed. In~other words, we start by searching for the extrema of the function
\begin{equation}
\label{eq:W_N_pdf}
\frac{1}{\sqrt{2\pi}}\exp\left({-\frac{(W(\gamma z))^2}{2\gamma^2}}\right)\frac{\exp({-W(\gamma z)})} {1+W(\gamma z)}.
\end{equation}

To investigate the existence of extrema for different values of $\gamma>0$, we first have to take the derivative from the expression \eqref{eq:W_N_pdf} by $z$, ignoring the constant in front:
\begin{equation}
\label{eq:W_N_deriv1}
\begin{aligned}
\left(\frac{\exp\left({-\frac{(W(\gamma z))^2}{2\gamma^2}-W(\gamma z)}\right)}{1+W(\gamma z)}\right)'
&=\frac{\exp\left({-\frac{(W(\gamma z))^2}{2\gamma^2}-W(\gamma z)}\right) \left({-\frac{(W(\gamma z))^2}{2\gamma^2}-W(\gamma z)}\right)'}{(1+W(\gamma z))} \\
&-\frac{\exp\left({-\frac{(W(\gamma z))^2}{2\gamma^2}-W(\gamma z)}\right)(1+W(\gamma z))'}{(1+W(\gamma z))^2}.
\end{aligned}
\end{equation}
{Using} 
 Formula \eqref{eq:Wprim}, we can~write
\begin{equation*}
\begin{aligned}
\left({-\frac{(W(\gamma z))^2}{2\gamma^2}-W(\gamma z)}\right)'
&=-\frac{2W(\gamma z)W'(\gamma z)}{2\gamma^2}-\gamma W'(\gamma z)  \\
&=-\frac{W(\gamma z)\exp{(-W(\gamma z))}}{\gamma(1+W(\gamma z))} -\frac{\gamma \exp{(-W(\gamma z))}}{1+W(\gamma z)} \\
&=\frac{-\exp{(-W(\gamma z))}(W(\gamma z)+\gamma^2)}{\gamma (1+W(\gamma z))}
\end{aligned}
\end{equation*}
and
\begin{equation*}
(1+W(\gamma z))'=\gamma W'(\gamma z)= \frac{\gamma \exp{(-W(\gamma z))}}{1+W(\gamma z)}.
\end{equation*}
{Now,} 
 substituting the results into Formula \eqref{eq:W_N_deriv1} leads to
\begin{adjustwidth}{-\extralength}{0cm}
\begin{equation}
\label{eq:W_N_pdf_deriv2}
\frac{\exp\left({-\frac{(W(\gamma z))^2}{2\gamma^2}-W(\gamma z)}\right)(-\exp(-W(\gamma z)))((W(\gamma z))^2+(1+\gamma^2)W(\gamma z)+2\gamma^2)}{\gamma(1+W(\gamma z))^3}=0.
\end{equation}
\end{adjustwidth}
{The} 
 equality \eqref{eq:W_N_pdf_deriv2} holds if the numerator is zero and the denominator is not. As we assume that $\gamma>0$, the~denominator provides a restriction $z\neq -\frac{1}{\gamma e}$, which is {already accounted for}. In~the numerator, because~the value of the exponential function is positive for any fixed argument (except for $z=0$ for the non-principal branch, which is dealt with separately), we need to solve the following quadratic equation:
\begin{equation}
\label{eq:W_N_Wquad}
(W(\gamma z))^2+(1+\gamma^2)W(\gamma z)+2\gamma^2=0
\end{equation}
with respect to $W(\gamma z)$.
The solution to this equation is of the form
\begin{equation}
\label{eq:W_N_Wquad_sol}
W(\gamma z)=-\frac{1+\gamma^2}{2}\pm \sqrt{\left(\frac{1+\gamma^2}{2}\right)^2-2\gamma^2} =\frac{-1-\gamma^2\pm\sqrt{\gamma^4-6\gamma^2+1}}{2}.
\end{equation}

{Let} 
 us now look more closely at the expression under the square root that determines the number of solutions for Formula \eqref{eq:W_N_Wquad_sol}.
Solving this equation $\gamma^4-6\gamma^2+1=0$ with respect to $\gamma$ results in
\begin{equation*}
\gamma^2=3\pm2\sqrt{2} \Leftrightarrow \gamma=\pm\sqrt{3\pm2\sqrt{2}}= \pm(\sqrt{2}\pm1).
\end{equation*}
{Because} 
 of the initial assumption $\gamma>0$, we are interested in two of these four solutions: $\gamma^{(1)}=\sqrt{2}-1\approx 0.4142$ and $\gamma^{(2)}=\sqrt{2}+1\approx 2.4142$.

We have established that the pdf of a Lambert $W\times N(0,1)$ random variable has two extrema in the following regions of the skewness parameter $\gamma$:  $\gamma>\sqrt{2}+1$  $0<\gamma<\sqrt{2}-1$. If~$\sqrt{2}-1<\gamma<\sqrt{2}+1$, then there are no real solutions for Formula \eqref{eq:W_N_Wquad_sol}. 

Now, let us restrict ourselves to the principal branch of the pdf and check which of the found values for $\gamma$ are within the range of values of the principal branch, i.e.,~$W(\gamma z)>-1$. Formula \eqref{eq:W_N_Wquad_sol} then leads to the~equality
\begin{equation*}
\frac{-1-\gamma^2\pm\sqrt{\gamma^4-6\gamma^2+1}}{2}>-1 \Leftrightarrow  \gamma^2-1<\pm\sqrt{\gamma^4-6\gamma^2+1},
\end{equation*}
where the solution for positive values for $\gamma$ is $0<\gamma\leq \sqrt{2}-1$. Thus, the~function has two extrema in the interval $0<\gamma\leq \sqrt{2}-1$ and is monotone decreasing for $\gamma>\sqrt{2}-1$. The~lemma is~proved.
\end{proof}

\begin{proof}[Proof of Lemma 3]
To prove this result, we can use the reasoning in the previous proof up to Formula \eqref{eq:W_N_Wquad_sol}, as~this holds for both the principal and non-principal branches. Then, it sufficient to check which solutions comply with the restriction $W(\gamma z)<-1$. Solving the {inequality}
\begin{equation*}
\frac{-1-\gamma^2\pm\sqrt{\gamma^4-6\gamma^2+1}}{2}<-1 \Leftrightarrow  \gamma^2-1>\pm\sqrt{\gamma^4-6\gamma^2+1}
\end{equation*}
for positive values of $\gamma$ results in $\gamma\geq \sqrt{2}+1$. 

Thus, the~function has two extrema if  $\gamma \in (\sqrt{2}+1,\infty)$, and~we have proved part (b) of the lemma. Similarly, the~Equation \eqref{eq:W_N_Wquad} has no real solutions for $\gamma<\sqrt{2}+1$. Now, the~assertion that $\lim_{z{\rightarrow} 0} f_{-1}(z)=0$ follows from the construction (see Formula \eqref{eq:fm1}), which proves part (a) of the lemma. The~lemma is~proved.
\end{proof}

\section[\appendixname~\thesection]{Details of Estimation}
\label{sec:app_estim}

We use several R \citep{viitaR} packages for parameter estimation.
For the hyperbolic, generalized hyperbolic, variance gamma, and normal inverse Gaussian distributions, we use a routine from the \emph{ghyp} package \citep{pakett-ghyp}; for the skew-normal and skew $t$ distributions, we use the~\emph{sn} package \citep{pakett-sn}; and for all other cases, we use the~\emph{fitdistrplus} package \citep{pakett-fitdistrplus}. In addition, we apply several functions from the \emph{LambertW} package \citep{Goerg2022pakett} to produce the pdf and cdf for the Lambert $W$ normal~distribution.

To access the US indemnity data, we use the R package \emph{fExtremes} \citep{pakett-fExtremes}, while for the Danish fire loss~data we use the \emph{copula} package  \citep{pakett-copula}.

We use the default starting values in each package for the relevant MLE routines, except~for the case of the Lambert $W$ distributions. {As the Lambert $W$ approach is relatively new, the~consistency and stability properties of the MLE estimator have not been thoroughly studied, though~the simulations provided in \citep{Goerg2011} are promising. In~the following,} we apply the method of moments to find the starting point for MLE. Next, we provide a more detailed overview of our selection of these starting~values.

\textbf{Lambert $W$ normal distribution.}{ To derive} 
 the starting values for the Lambert $W\times N(\mu ,\sigma)$ distribution{,} we use the mean, variance, and skewness coefficient of the Lambert $W\times N(\mu ,\sigma)$ random variable $Y$ provided {in} Formulas \eqref{eq:WnormEY}--\eqref{eq:Wnorm-skewness}.  
We first equate \eqref{eq:Wnorm-skewness} with the sample skewness coefficient and solve it numerically to produce $\gamma_0$. Next, the~expressions for the mean and variance are used, first substituting $\gamma_0$ and a sample variance $s_y^2$ of $Y$ into \eqref{eq:WnormDY} and~then solving it for $\sigma$ to obtain the starting~value of
\begin{equation*}
\sigma_{0} = \sqrt{\frac{s_y^2}{(e^{\gamma_0^2}(e^{\gamma_0^2}(1+4\gamma_0^2)-\gamma_0^2))}}.
\end{equation*}
{Lastly,} 
we substitute $\gamma_0$, $\sigma_0$ and the sample mean $\bar y$ into \eqref{eq:WnormEY} and solve for $\mu$ to~obtain

\begin{equation*}
\mu_{0}=\bar{y} - {\sigma_{0}} \gamma_0 e^{\gamma_0^2/2}.
\end{equation*}

\textbf{Lambert $W$ exponential distribution.} In the case of the Lambert $W\times Exp(\lambda)$ distribution, we use  the formula for the skewness coefficient \eqref{eq:WxEgamma1} with the sample-based estimate $\hat\gamma_1$ to find a starting value $\gamma_0$ for the skewness parameter.
As {the} skewness coefficient $\gamma_1$ is a non-monotone function of $\gamma$ (see Figure~\ref{fig:WExp-skewness}), only solutions in the interval $(-1, \frac13)$ are used, as values $\gamma < -1$  produce too drastic of truncation. For~the rate parameter $\lambda$, we use the expression of the first moment and solve  $\bar y = \frac{1}{\lambda(1-\gamma_0)^2}$ 
to derive the formula 
\[
\lambda_0 = \frac{1}{\bar y(1-\gamma_0)^2}
\]
for {the} starting value of $\lambda$.

\section[\appendixname~\thesection]{Data Histograms  with the Three Best Fitting Models}
\label{sec:app_joonised}

\begin{figure}[H]
\includegraphics[width=0.9\textwidth]{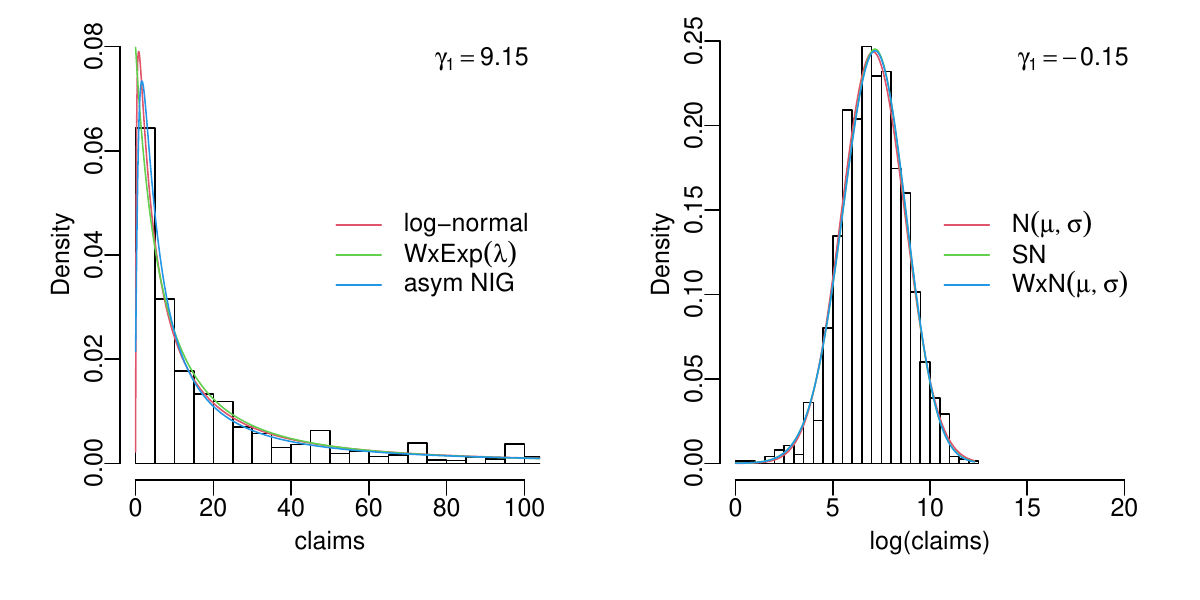}
\caption{Left panel: US indemnity data (in thousands of USD). {For a better overview,} values above 100 are not shown on the histogram. Right panel: the same data after log-transformation. The added lines represent the best three estimates based on BIC.}
\label{fig:data-usa-3joont}
\end{figure}
\unskip

\begin{figure}[H]
\includegraphics[width=0.9\textwidth]{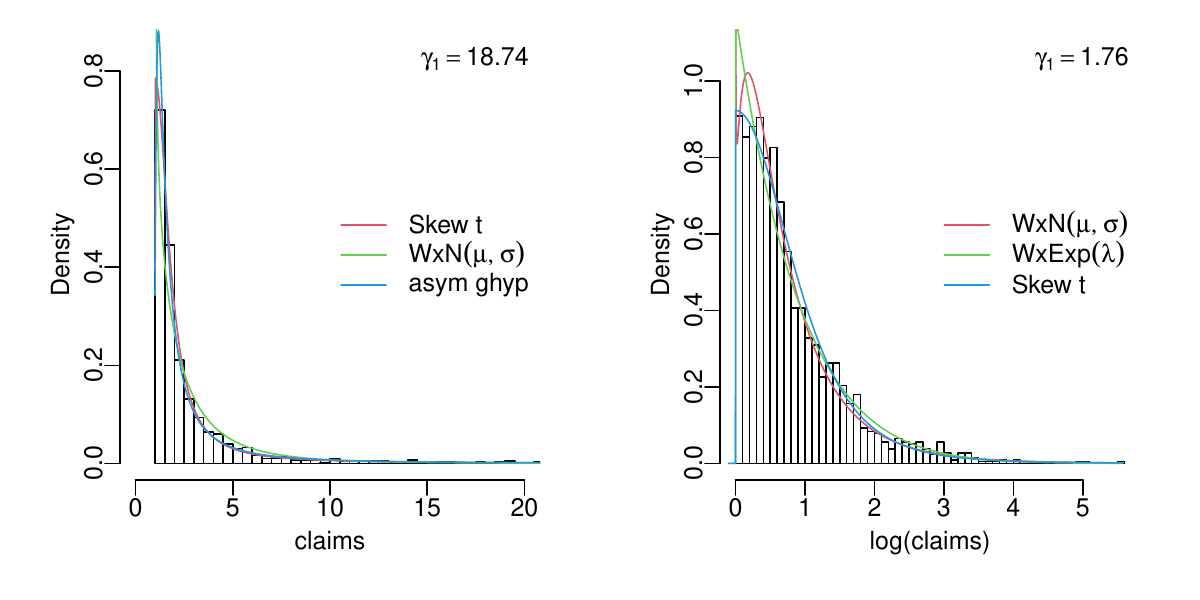}
\caption{Left panel: Danish fire data claims (in millions of DKK). {For a better overview,} values above 20 are not shown on the histogram. Right panel: the same data after log-transformation. The added lines represent the best three estimates based on BIC.}
\label{fig:data-taani-3joont}
\end{figure}

\begin{adjustwidth}{-\extralength}{0cm}

\reftitle{References}

\PublishersNote{}
\end{adjustwidth}
\end{document}